\documentclass[10pt]{iopart} % 12 pt if not two column

\usepackage{graphicx}% Include figure files
\usepackage{hyperref}% add hypertext capabilities
\usepackage[dvipsnames]{xcolor}
\usepackage{siunitx}
\usepackage{braket}
\usepackage{placeins}
\usepackage[noadjust]{cite}
\usepackage{float}
\usepackage{parskip}
\usepackage[utf8]{inputenc}

\begin{document}

\title{Photon correlation transients in a weakly blockaded Rydberg ensemble}

\author{Charles M\"{o}hl$^{1,2}$, Nicholas L. R. Spong$^1$, 
	Yuechun Jiao$^{1,3}$, Chloe So$^1$, Teodora Ilieva$^1$, Matthias Weidem\"{u}ller$^2$ and Charles S. Adams$^1$}

\address{$^1$ Joint Quantum Centre (Durham-Newcastle), Department of Physics, Durham University, South Road, Durham, DH1 3LE, United Kingdom}
\address{$^2$ Physikalisches Institut, Universität Heidelberg, Im Neuenheimer Feld 226, 69120 Heidelberg, Germany}
\address{$^3$ State Key Laboratory of Quantum Optics and Quantum Optics Devices, Institute of Laser Spectroscopy, Shanxi University, Taiyuan 030006, China}
\ead{c.s.adams@dur.ac.uk}
\vspace{10pt}
\begin{indented}
\item[]October 2019
\end{indented}

\begin{abstract}
    The non-linear and non-local effects in atomic Rydberg media under electromagnetically induced transparency (EIT) make it a versatile platform for fundamental studies and applications in quantum information. In this paper, we study the dynamics of a Rydberg-EIT system in an ensemble that allows for more than one Rydberg excitation in the propagation direction. The density of two-level atoms is such that transient superradiant effects occur. We experimentally observe a cross-over between coherent collective emission (`flash') of two-level atoms to a Rydberg dressed regime ('dressed flash') under EIT condition. The complex dynamics are characterised using both intensity and time correlation measurements. We show that while steady-state EIT gives a second order correlation $g^{(2)}=0.79\pm 0.04$, the Rydberg-dressed flash exhibits anti-bunching down to $0.2\pm0.04$.
\end{abstract}

\vspace{2pc}
\noindent{\it Keywords}: Rydberg blockade, Collective response, Single photon\\

\vspace{2pc}
\submitto{\JPB}

%\maketitle
\ioptwocol

%%%%%%%%%%%%%%%%%%%%%%%%%%%%%%%%
%%%% Begin of the main text %%%%
%%%%%%%%%%%%%%%%%%%%%%%%%%%%%%%%

\section{\label{sec:intro}Introduction}

    First demonstrated in 2007 \cite{Moh2007}, electromagnetically-induced transparency (EIT) involving highly-excited Rydberg states (with principal quantum number $n>60$) creates a medium with an extreme optical non-linearity \cite{Pri2010} that enable strong photon-photon interactions \cite{Peyronel2012} and all-optical photon gates \cite{Tiarks2019}. The closely related four-wave mixing combined with Rydberg interactions have been used to generate non-classical light \cite{Dudin2012,Ripka2018}, offering a promising single-photon source. The interplay of excitation, interactions and light propagation is a complex many-body problem \cite{Gorshkov2011,Gorshkov2013,Khazali15,Sevincli2011,gorniaczyk16,petrosyan11,Li14,Firstenberg2013}.
    
    One aspect that is often overlooked is that light-matter interactions in dense atomic ensembles exhibit collective effects \cite{Jenkins2016,Roof2016,Guerin2016,Bradac2017,Facchinetti2016a,Guerin2017,Chen2010a,Toyoda1997} such as superradiance, however the interplay between such collective effects and Rydberg interactions is not well understood. For this reason, in this paper we study the crossover between coherent collective emission (`flash') from a dense two-level atom ensemble \cite{Bettles2018}, and single-photon emission from an ensemble of three-level atoms where the excited state of the two-level system is coupled to a highly-excited Rydberg state. While similar to previous studies on interferecene effects in collective ensembles of two- and three-level systems \cite{Chen2010}, in this work transient emission with sub-poissonian statistics is observed after the EIT steady state. We refer to this case as a Rydberg dressed `flash'.
    
    The transient and steady-state EIT optical response of a dense Rydberg atomic gas are investigated using both intensity and time correlation measurements. The results obtained further highlight the importance of interaction-induced dephasing even beyond the blockade radius, and demonstrate how the transient collective response can be exploited to generate strongly anti-bunched light pulses despite imperfect blockade of the medium.

\section{\label{sec:experiment}Experiment}
 \begin{figure*}%[!htb]
	\centering
	\def\svgwidth{7cm}
	\includegraphics[width=6.162in]{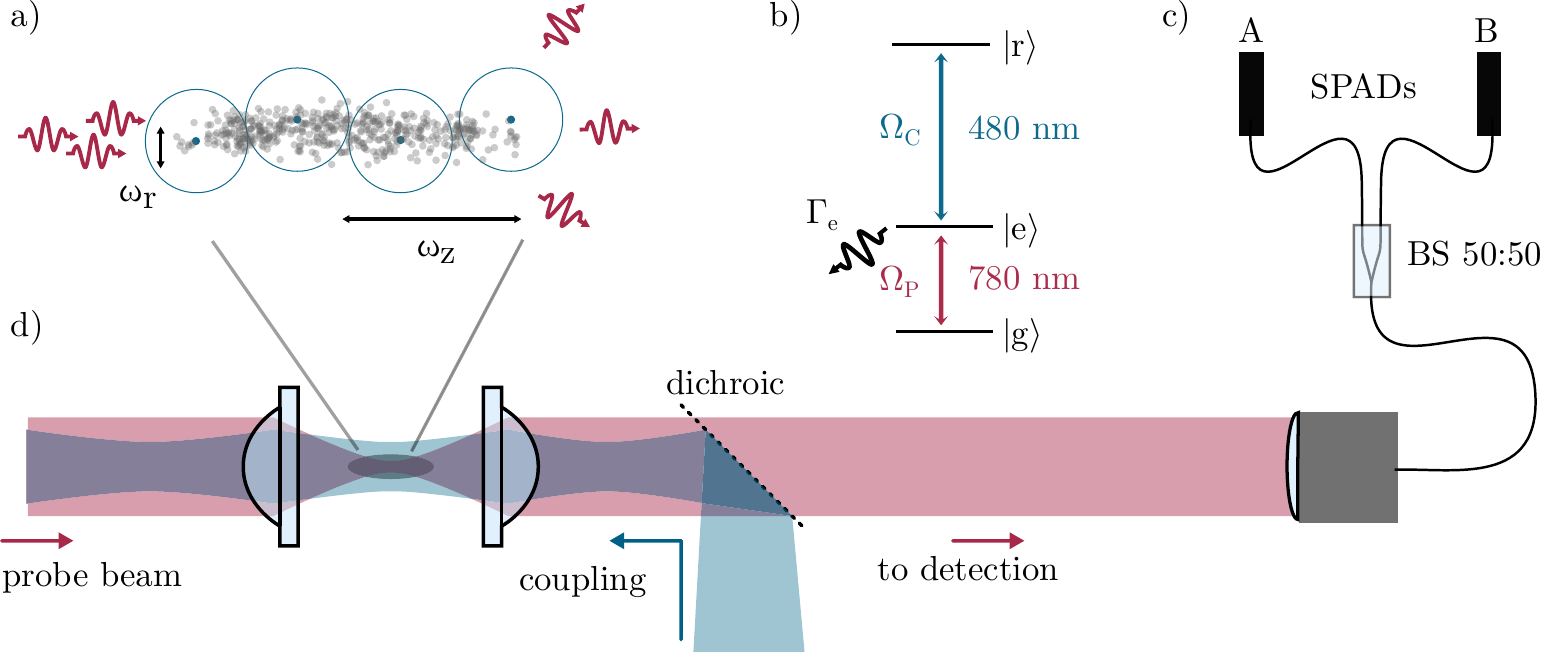}
	\caption{\label{fig:setup}%
		Schematic of the experimental scenario: (a) A cold and dense cigar-shaped cloud of $^{87}$Rb atoms (grey dots) is probed using a weak laser pulse on resonance with the probe transition ($\ket{5S_{1/2},F=2,m_\text{F}=2}\longrightarrow \ket{5P_{3/2},F'=3,m_\text{F'}=3}$, denoted $\Omega_\text{P}$ in (b)) and coupled to a high Rydberg state $\ket{r}=\ket{nS_{1/2}}$ in an EIT ladder scheme using a strong CW coupling laser ($\Omega_\text{C}$ in (b)). (d) The cold atomic cloud is prepared by magneto-optical and successive dipole trapping between two high-NA lenses, the first one being used to focus the probe and dipole trap beams to a waist of $w_\text{P}\approx\SI{1.4}{\micro\metre}$ and $w_\text{r}\approx\SI{4.5}{\micro\metre}$ respectively. The counterpropagating coupling beam has a waist of $w_\text{C}\approx\SI{25}{\micro\metre}$. The second lens collects the fluorescence, which is then detected downstream on a HBT interferometer setup using single-photon avalanche detectors 
		(c).
	}%
\end{figure*}

    %\textcolor{blue}{Figure 1}
    The details of the experimental setup are described elsewhere \cite{Busche2016}. We prepare on the order of $1000$ $^{87}$Rb atoms in a dipole trap with a radial waist $w_\text{r}=\SI{4.5}{\micro\metre}$ and longitudinal spread along the beam axis of approximately $w_\text{z}=\SI{60}{\micro\metre}$ (Figure \ref{fig:setup}a). The atoms can either be excited to the intermediate excited state $\ket{e}$ or to a Rydberg state $\ket{r}=\ket{nL}$, where $n$ is the principal quantum number ($n=80$ in the results reported below), and $L$ is the orbital angular momentum. The long-range dipole-dipole interactions (Van-der-Waals or resonant dipole-dipole) limit a second Rydberg excitation within a characteristic blockade radius, thereby forming a Rydberg superatom \cite{Weber2015} (shown as blue circles in Figure~\ref{fig:setup}a) with long range spatial correlations \cite{Zeuthen2017a}.
    \par
    The three-level EIT ladder system (Figure \ref{fig:setup}b) consists of the D2-line probe transition ($\Omega_\text{P}$) at $\SI{780}{\nano\metre}$ between ground state $\ket{g}$ and excited (intermediate) state $\ket{e}$, and the coupling transition ($\Omega_\text{C}$) at $\SI{480}{\nano\metre}$ between $\ket{e}$ and Rydberg state $\ket{r}$. The atomic cloud is illuminated in a counter-propagating fashion with probe (red) and coupling laser beams (blue) as shown in Figure \ref{fig:setup}d. The probe beam is tightly focused ($w_\text{P}\approx\SI{1.4}{\micro\metre}$) by a high numerical aperture (NA) lens and its transmission (T) is collected downstream by a single mode (SM) fiber used as a spatial filter, and detected on a standard Hanbury Brown and Twiss (HBT) interferometer \cite{Busche2016} comprising a fiber beam splitter and two single-photon avalanche detectors (SPADs) (Figure \ref{fig:setup}c).
    The intensity measurements shown in this work have a binning size of $\SI{5}{\ns}$ and are normalised to the reference intensity $I_0$, which represents the number of detection events per bin and shot from the probe pulse without atoms present.
    
    \begin{figure*}
    	\centering
    	\includegraphics[width=\textwidth]{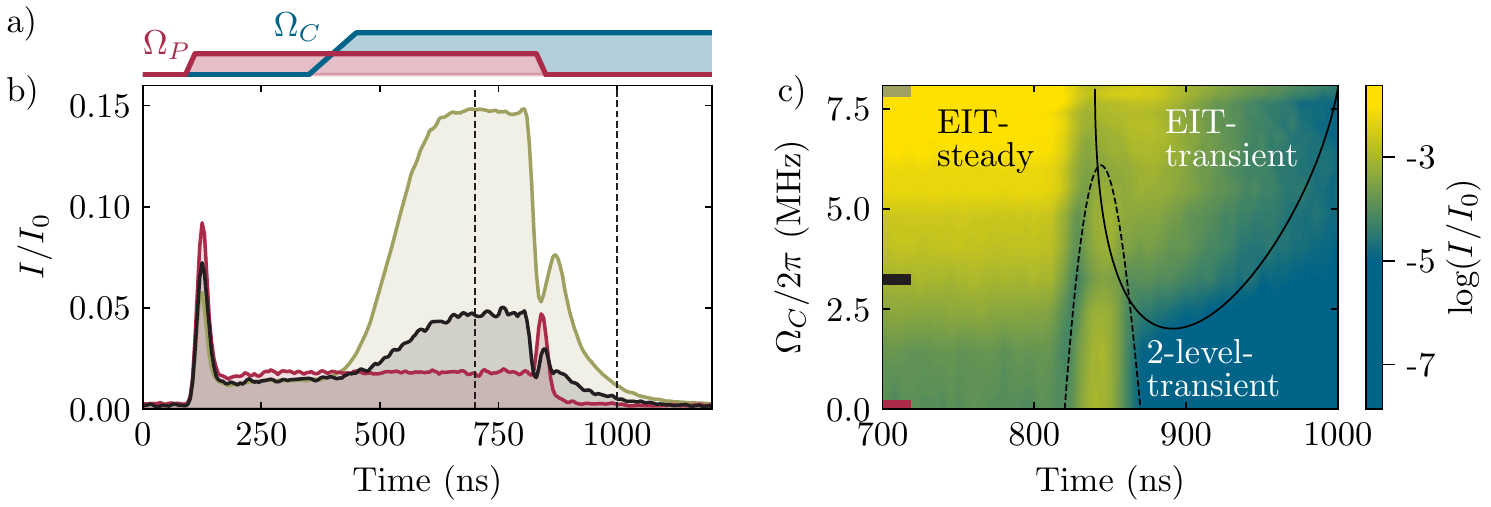}
    	\caption{\label{fig:arrivals}%
    	    Crossover from collective 2-level transient to Rydberg dressed EIT-transient for $\ket{nL}=\ket{80S_{1/2}}$: (a) Schematic of the laser pulse sequence used for the data of this Figure. (b) Normalised transmitted intensity as a function of time for $\Omega_\text{C} = 0$ (red), for the intermediate coupling regime $\Omega_\text{C} = 2\pi\cdot\SI{3.24\pm0.10}{\mega\hertz}$ (black) and for high coupling laser power $\Omega_\text{C} = 2\pi\cdot\SI{8.08\pm0.24}{\mega\hertz}$  (olive). (c) Contour plot of normalised transmission on two-photon resonance for different coupling powers $\Omega_\text{C}$. $2.5\times10^6$ pulse sequences per trace were recorded for the normalised intensity data shown here, at a reference intensity $I_0$ of $7\cdot10^{-3}$ counts per shot and time-bin ($\SI{5}{\ns}$). 
    	}%
    \end{figure*}

    %\textcolor{blue}{Figure 2}
    \section{Results}
    %\subsection{Transient response of the two-level and EIT-system}
    The laser pulse sequence used for the characterisation of the optical response of the medium is shown in Figure \ref{fig:arrivals}a. An on-resonance weak coherent probe pulse ($\Omega_\text{P}$) containing on the order of $5$ photons with a duration of $\SI{700}{\nano\second}$ and rise/fall time $\SI{20}{\nano\second}$ is followed by a strong coupling laser pulse ($\Omega_\text{C}$) with a delay of approximately $\SI{350}{\nano\second}$ and rise/fall time of $\SI{100}{\nano\second}$. The coupling beam delay is implemented to monitor the optical depth (OD) by comparing the relative photon counts from the optical response in the two-level system steady state (between $\SI{200}{\nano\second}$ and $\SI{350}{ns}$) to that of the probe laser pulse in the absence of atoms, and to investigate possible EIT transients, see e.g. \cite{zhang_bai_huang_2018}. The normalised transmitted intensity through the medium for three different coupling powers ($\Omega_\text{C}=2\pi\cdot\SI{0}{\mega\hertz}$, $\Omega_\text{C} = 2\pi\cdot\SI{3.24\pm0.10}{\mega\hertz}$ and $\Omega_\text{C} = 2\pi\cdot\SI{8.08\pm0.24}{\mega\hertz}$) are shown in Figure \ref{fig:arrivals}b.
    
    In general, the observed transmission is a consequence of the interference between the incident laser field and the scattered dipole emission. For the two-level case ($\Omega_\text{C}=2\pi\cdot\SI{0}{\mega\hertz}$; in red) it shows a distinct first peak at the rising edge of the probe pulse, followed by a steady state with low photon counts determined by the optical depth of the medium. A second peak appears at the falling edge of the pulse as a result of the interference of the decaying collective modes of the atomic polarization \cite{Bettles2018}. As discussed below, this two-level transient has a superradiant behavior with collectively enhanced decay rate of $\Gamma_1=(2.3\pm0.1)\Gamma_0$ in this experiment.
    
    The time dependent intensity in the intermediate regime of the crossover between the collective two-level transient and the EIT transient or (Rydberg dressed) 'flash' is shown in black (Figure \ref{fig:arrivals}b.  The adressed Rydberg state here is $\ket{nL}= \ket{80S_{1/2}}$ with a coupling Rabi frequency $\Omega_\text{C} = 2\pi\cdot\SI{3.24\pm0.10}{\mega\hertz}$. The build-up of the EIT window manifests as an increase in photon counts from $\SI{500}{\nano\second}$ on.
    The height of the coherent two-level transient ('flash'), with decay rate $\Gamma_1$, at the falling edge is reduced and a second slow decaying feature, with  decay  rate $\Gamma_2$, appears thereafter. This EIT-transient or 'dressed flash' dominates the dynamics for higher coupling Rabi frequencies, as shown for $\Omega_\text{C} = 2\pi\cdot\SI{8.08\pm0.24}{\mega\hertz}$ (in olive) and reaches a peak height of well over $\SI{5}{\percent }$ of the incoming pulse intensity.
    Note, that the intensity in the EIT steady state is below $\SI{15}{\percent}$ indicating imperfect transparency compared to single particle prediction. We will later motivate the relation to interaction induced dephasing, which leads to a broadening of the EIT feature and by that decreases the EIT-transient times as discussed by Zhan et al. \cite{zhang_bai_huang_2018}.
    In contrast however to their theoretical calculations, where oscillatory behaviour of the transient is expected under certain conditions, the decay of the Rydberg excitations seems to be fully dominated by interaction-induced dephasing, leading to an exponential decay without resolvable oscillations \cite{Moehl2019}.
    
    The continuous transition from two-level to EIT-transient is depicted in the contour plot in Figure \ref{fig:arrivals}c.
    It shows a considerable group delay of the EIT transient of $\SI{28}{\nano \second}$ compared to the two-level transient, which indicates that dispersive effects like the slow light effect also play a role in this ensemble \cite{Peyronel2012}.
    
    \begin{figure}
    	\centering
    	\includegraphics[width=0.9\columnwidth]{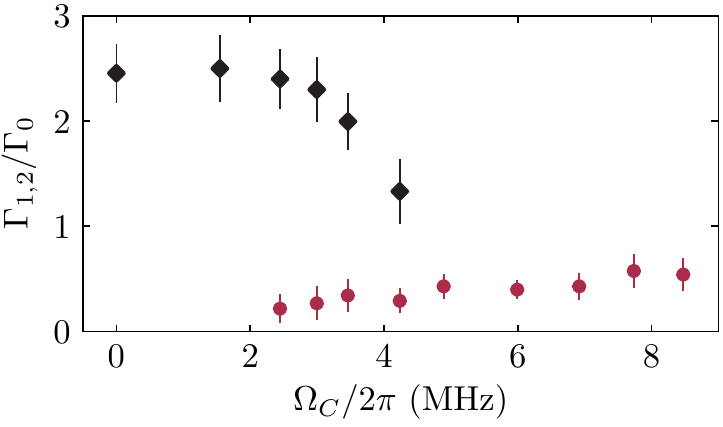}
    	\caption{\label{fig:decays}%
    		Transient decays for different coupling strengths $\Omega_\text{C}$: Decay rates $\Gamma_1$ and $\Gamma_2$ of the collective 2-level and EIT-transient respectively as a function of coupling strength $\Omega_\text{C}$ in squares (black) and circles (red) respectively for Rydberg state $\ket{nL}=\ket{80S_{1/2}}$.
    	}%
    \end{figure}
    
    %\textcolor{blue}{Figure 3}
    The transient dynamics are further investigated by considering the decay rates of the two-level ($\Gamma_1$) and EIT-transient ($\Gamma_2$) for different coupling powers, as shown in Figure \ref{fig:decays}. Here, the decay rates are normalized to the natural line width of the D$_2$ transition $\Gamma_0=2\pi\cdot\SI{6.065}{\mega\hertz}$ \cite{Steck}. All error bars correspond to the uncertainty of the exponential fits to the transients.
    The coherent decay, $\Gamma_1$, starts at around $2.3\Gamma_0$, in agreement with the expected collective speed-up at OD$=4$ \cite{Bettles2018}, and tends towards $\Gamma_0$ for increasing $\Omega_\text{C}$, indicating a breakdown of the collective two-level response by the Rydberg-blockade mechanism.
    The dressed decay, $\Gamma_2$, starts at around $0.2\Gamma_0$ and increases slightly to $0.5\Gamma_0$ with increasing coupling strength. Note that, due to limited coupling laser power, higher Rabi frequencies were not accessible for $\ket{r}=\ket{80S_{1/2}}$.
    We propose to explain the increase of the EIT-transient decay rate $\Gamma_2$ with coupling power with a weakened blockade mechanism compared to interaction induced dephasing. This unites different pictorial descriptions of Rydberg-Rydberg interactions in EIT media: With increasing $\Omega_\text{C}$ the blockade radius $r_\text{b}\propto\Omega_\text{C}^{-1/6}$ is reduced, allowing smaller distances and therefore stronger interactions and dephasing between Rydberg excitations, before they blockade each other. This reduction of the blockade mechanism is also suggest from the increase in the EIT-steady state intensity at high compared to low $\Omega_\text{C}$ in Figure \ref{fig:arrivals}.
    
    \par
    \begin{figure*}%[h!]
    	\centering
    	\includegraphics[width=0.95\textwidth]{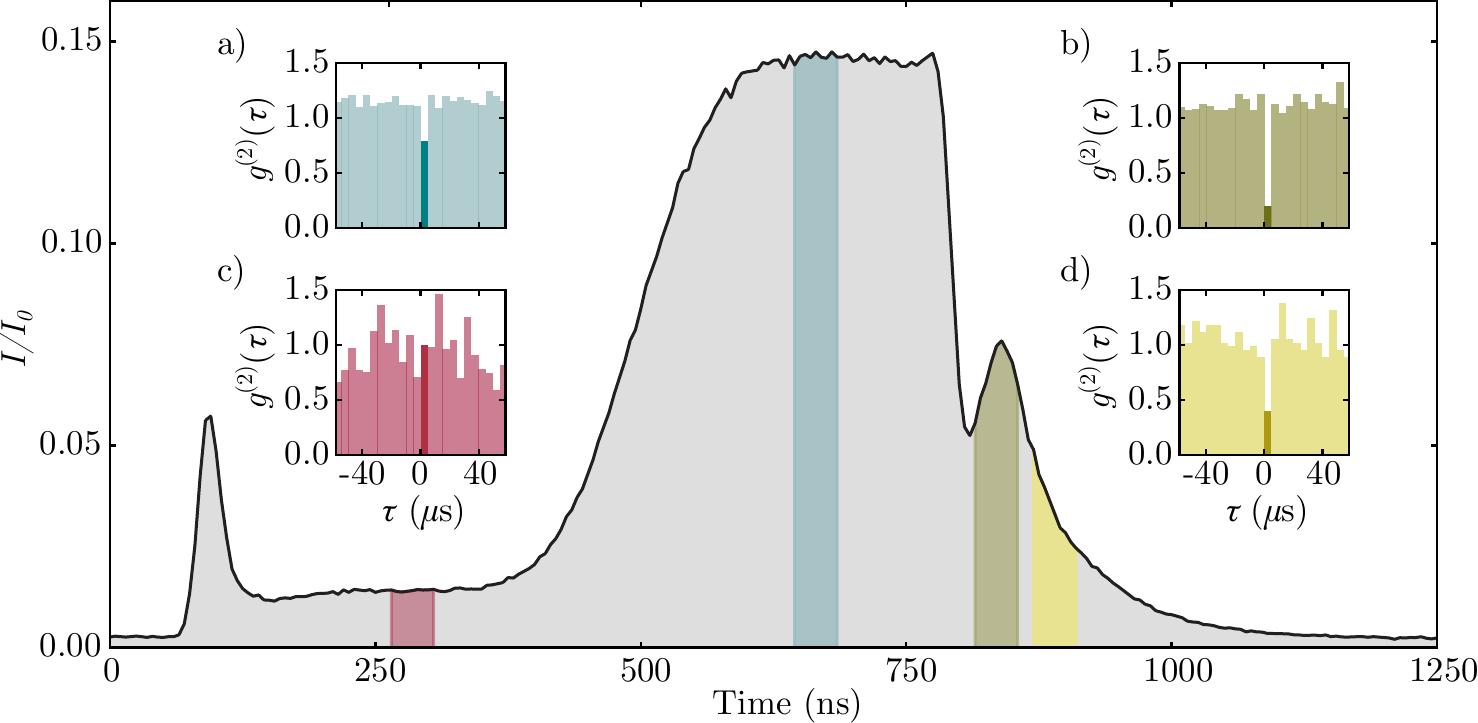}
    	\caption{\label{fig:dflash}%
    	Detailed analysis of the photon statistics within the pulse sequence at $\ket{nl}=\ket{80s_{1/2}}$: Normalised transmitted intensity versus time and second-order autocorrelation function $g^{(2)}$ through a cloud of (c) opaque two-level atoms with $g^{(2)}(0)=1.00\pm0.12$, (a) three-level atoms under EIT condition with $g^{(2)}(0)=0.79\pm0.04$, (b) first $\SI{50}{\ns}$ of the EIT-transient with $g^{(2)}(0)=0.20\pm0.04$, (d) second $\SI{50}{\ns}$ of the EIT dressed flash with $g^{(2)}(0)=0.40\pm0.11$. The uncorrelated levels for $g^{(2)}(\tau\neq0)$ are $0.94$, $1.16$, $1.13$ and $1.07$ respectively for the four cases. $17.5\times10^6$ pulse sequences were recorded for the normalised intensity and correlation data shown here, at a reference intensity $I_0$ of $8\cdot10^{-3}$ counts per shot and time-bin ($\SI{5}{\ns}$). The time-bin for the correlation function $g^{2}$ corresponds to the total sequence length of $\SI{4935}{\ns}$.
    	}%
    \end{figure*}
    
    %\textcolor{blue}{Figure 4}
    %\subsection{Transient photon correlations}
    The quantum nature of the dynamics in our system is revealed in the statistical correlations of the transmitted photons within different parts of the sequence as shown in Figure \ref{fig:dflash} for coupling Rabi frequency $\Omega_\text{C} = 2\pi\cdot\SI{8.1\pm0.2}{\mega\hertz}$. As expected from a two-level system driven by a coherent state (incoming laser pulse) the statistics show no suppression of photon coincidences (red, $\SI{150}{\ns}$ to $\SI{350}{\ns}$). In the steady state response of the three-level system under Rydberg-EIT condition (blue, $\SI{570}{\ns}$ to $\SI{770}{\ns}$) a suppression of detected coincidences  \cite{Weber2015} with $g^{(2)}(0)=0.79\pm0.04$ is observed. This suppression stems from the Rydberg blockade effect, which causes the incoming probe photons to be localized in the medium as Rydberg-polaritons and propagate through the medium at reduced group velocity as a result of the slow light effect found in EIT media \cite{Hau1999}. Furthermore, the Rydberg-blockade imposes a minimum spacing between the polaritons, which translates into a temporal separation of the transmitted photons \cite{Zeuthen2017a}, leading to anti-bunching. As a consequence of the rather low optical depth per blockade radius OD$_\text{b}$, which is estimated to be around $0.4$, the effect of the Rydberg-blockade on incident photons is expected to be reduced, compared to other experiments in regimes with OD$_\text{b}>>1$ \cite{Peyronel2012}.
    
    Interestingly however despite the low OD$_b<1$, a pronounced anti-bunching of photons is observed in the EIT-transient. The time window centered around the maximum of the feature (olive) shows the strongest supression of coincidences with $g^{(2)}(0)=0.20\pm0.04$. The imperfection of coincidence suppression mainly stems from the elongated cigar-shape of the atomic ensemble, allowing more than one Rydberg excitation to be simultaneously present in the medium \cite{Maxwell2013}. For the Rydberg state $\ket{80S_{1/2}}$ the blockade radius, $r_\text{b}$, is approximately $\SI{10}{\micro\meter}$, allowing for roughly two simultaneous excitations until the cloud is fully blockaded. As will be discussed in more detail later, the higher $g^{(2)}(0)$ value at later times within the transient (see Figure \ref{fig:dflash}d), suggests complex correlation dynamics in the system. These dynamics seem different in nature to those found in storage protocols as shown in \cite{Moehl2019}.
    It should be noted, that due to narrow width and low intensity, the initial two-photon transient pulse at around $\SI{100}{\ns}$ did not provide sufficient counts to conclusively determine $g^{(2)}$. The time dependent intensity measurements are limited by the systematic stability of the experiment over the course of data collection (~hours).
    
    The anti-bunching effect increases with principal quantum number $n$ (see \cite{Moehl2019}). The obtained results are expected due to the strong scaling of the blockade radius with $n$ which defines the maximum number of excitations within the atomic cloud, as well as the Rydberg-Rydberg interactions dephasing \cite{Bariani2012}. Both the $g^{(2)}(0)$ value, and the photon rate increase with incident photon number (see \cite{Moehl2019}).

\begin{figure}
    	\centering
    	\includegraphics[width=1\columnwidth]{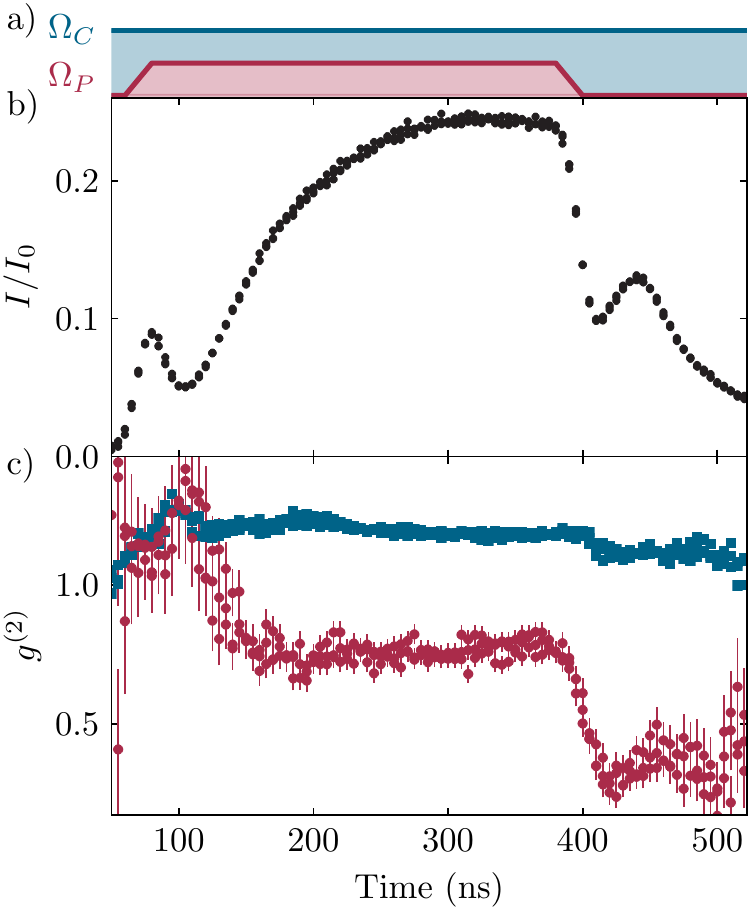}
    	\caption{\label{fig:correlations}%
    	Photon correlation transients in Rydberg-EIT media:
    	(a) Pulse sequence for the data shown in this figure. (b) Normalised intensity and (c) Second-order autocorrelation function $g^{(2)}(\tau)$ for $\tau=0$ (red) and the statistically averaged value for $\tau\neq0$ (blue). The reference intensity $I_0$ is $6\cdot10^{-3}$ counts per shot and time-bin ($\SI{5}{\ns}$). Data points shown are for three subsequent identical pulses in one experimental run, with a binning of $\SI{30}{\ns}$ for the correlation data in (c). $32.5\times10^6$ pulse sequences were recorded for the normalised intensity and correlation data shown here.
    	}%
    \end{figure}

    %\textcolor{blue}{Figure 5}
    In order to gain more information about the interplay between dephasing and the Rydberg-blockade, the transient intensity and correlations dynamics were analysed in more detail with a slightly modified pulse sequence shown in Figure \ref{fig:correlations}a. The coupling beam is operated in continuous wave (CW) mode with a Rabi frequency of $\Omega_\text{C} = 2\pi\cdot\SI{9.8\pm0.3}{\mega\hertz}$, while a weak probe pulse is sent through the medium. The pulse sequence is chosen to avoid any limitations due to the slower rising time of the coupling power. This leads to an almost symmetric shape of the EIT turn-on transient and decaying transient at the beginning and end of the probe pulse respectively (Figure \ref{fig:correlations}b.
    The calculated second-order autocorrelation function $g^{(2)}(\tau)$ for $\tau=0$ (red) and the statistically averaged value for $\tau\neq0$ (blue) for three subsequent identical pulses in one experimental run are presented in Figure \ref{fig:correlations}c, where a time binning of $\SI{30}{\ns}$ was used. In parallel to the build of EIT transparency, the transmitted photons exhibit start anti-bunching in the EIT steady state. The lowest $g^{(2)}(0)$ is observed in beginning of the EIT-transient, while the correlations show non-trivial behaviour in later parts of the feature.
    The non-uniform suppresion of two-photon events within the transient suggest that the retrieval dynamics of Rydberg excitations in the medium are different for two (or multiple) compared to a single excitation. In that sense, the regime of 'weak blockade' investigated here requires a first-order correction to the usually described (zero-order) blockade model \cite{Zeuthen2017a}, by including the presence of two simultaneous excitations that dephase each other.
    The equal-time correlation measurements $g^{(2)}(0)$ allow to resolve to some extend the contributions of single and multiple excitations to the observed intensity, corroborating that interaction-induced dephasing dominates the dynamics in the transient.
    Note, that the slight reduction of the $g^{(2)}(\tau\neq0)$ (blue) may be explained from the absence of the $\SI{2}{\percent}$ directly transmitted intensity from the probe pulse at OD $4$.
    
\begin{figure}
    \centering
    \includegraphics[width=1\columnwidth]{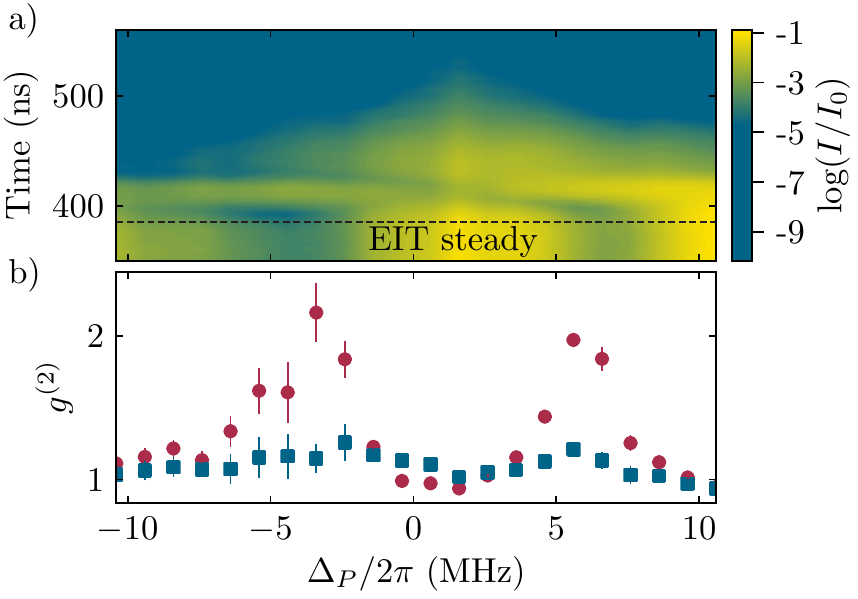}
    \caption{\label{fig:detuning}%
    	Detuning of the probe pulse ($\Delta_\text{P}\neq0$, $\Delta_\text{C}=0$): (a) Contour plot of the normalised intensity of the EIT steady state ($<\SI{385}{\ns}$) and EIT-transient ($>\SI{385}{\ns}$) for different probe detunings. (b) second-order autocorrelation function $g^{(2)}(\tau)$ for $\tau=0$ (red) and the statistically averaged value for $\tau\neq0$ (blue) in the EIT steady state. $1.5\times10^6$ pulse sequences per detuning were recorded for the normalised intensity and correlation data shown here. The time-bin for the correlation function $g^{2}$ in (b) corresponds to the total sequence length of $\SI{4440}{\ns}$.
    	}%
    \end{figure}

    %\textcolor{blue}{Figure 6}
    As previously discussed, the transients of the on-resonance EIT Rydberg excitation ($\Delta_\text{P/C} = 0$) depends on the dephasing induced by the Rydberg nature of the metastable third state $\ket{r}$ that lead to imperfect transparency. If $\ket{r}$ is a non-interacting narrow-linewidth state, the transparency window would lead to negligible absorption on two-photon resonance and make the process very slow. The idea now is to break the EIT-condition and excite to the Rydberg state via one of the dressed states $\ket{\pm} = 1/\sqrt{2}(\ket{r}\pm\ket{e})$, which might be described as a Rydberg-dressing of the two-level transient. The detuning should therefore be matched to the coupling Rabi splitting $\Delta_\text{P} \approx \pm \Omega_\text{C}/2$. For the data presented in Figure \ref{fig:detuning}, the probe laser detuning was varied for different spectral positions in the EIT spectrum. Similarly to before, the transient response behavior of the Rydberg-EIT as a function of the probe-field is then studied by collecting the emission from the intermediate excited state $\ket{e}$. The counter plot of the dressed flash intensity for different detunings (Figure \ref{fig:detuning}a shows that the transient response of the medium for red- and blue-detuning is strongly altered compared to the on-resonant case.
    The appearance of the two-level transient for $\Delta_\text{P}\neq0$ in the time resolved intensity suggests an analogy between detuning the probe beam from EIT resonance and reducing the coupling Rabi frequency $\Omega_\text{C}$ as shown previously.
    
    When looking at the photon correlations however, the distinction between the two becomes apparent.
    The second-order auto correlation function $g^{(2)}(\tau)$  for $\tau =  0$  (red)  and  the  statistically  averaged value for $\tau \neq 0$ (blue) were evaluated within the EIT steady state window between $\SI{245}{\ns}$ and $\SI{385}{\ns}$ (shown in Figure \ref{fig:detuning}b). Significant photon bunching in the transmitted light is observed for probe detunings matching the EIT splitting $\Delta_P\approx\pm\Omega_\text{C}/2$. Interestingly, this bunching appears for both red and blue detuning, which is in contrast to the usually described 'Rydberg enabling' of multiple excitations, when detuning the coupling beam and leaving the probe beam on resonance ($\Delta_\text{P}=0$, $\Delta_\text{C}\neq0$). The results are however in total agreement with predictions by \cite{Yan2019}, who attribute the bunching feature at the Autler-Townes peaks to dephasing mechanisms.
    The error bars in Figure 6b) are dominated by the relative error of the coincidence count. Therefore, the asymmetry of the error bars w.r.t. the detuning can be explained by the lower intensity at negative detunings.
    It should be noted, that the less pronounced anti-bunching feature on resonance here might stem from a slightly higher input photon number. As discussed in \cite{Moehl2019}, the anti-bunching effect decreases with increasing input photon rate.
    
    \section{\label{sec:conc}Conclusion} %%%%
    \par
    In conclusion, we have presented and characterised transient collective effects in a dense Rydberg-EIT ensemble of cold $^{87}$Rb atoms. The observed decay rates of the transients in EIT on-resonance regime suggests a transition from blockade-dominated dynamics at low coupling Rabi frequency $\Omega_\text{C}$, to increased interaction-induced dephasing at higher $\Omega_\text{C}$. Off-resonant excitation via the dressed states was also investigated, revealing significant photon bunching in the transmitted light within the EIT steady state in stark contrast to the bunching in the on-resonant case.
    The surprising fact that the weakly anti-bunched emission in the EIT steady state is followed by a stronger anti-bunched 'dressed flash' shows how single photon pulses can be created by exploiting collective interference effects of a merely weakly blockaded ensemble, in contrast to write \& read pulse sequences usually used in this context \cite{Dudin2012,Ripka2018,Maxwell2013}.
    From a fundamental point of view, our results unify the effective description of Rydberg interactions in the frequency domain and the direct consequences on the steady state and transient dynamics in the time domain. Specifically, the observed transients may be related to interaction-induced dephasing, while the non-uniform coincidence rate within the EIT-transient suggest that photon pairs tend to arrive at different times than single photons.
    We further demonstrated how the photon statistics in the EIT steady-state can be continuously tuned from anti-bunched to bunched light emission using the coupling laser detuning.
    By careful tuning of the experimental parameters like ensemble size and atomic density as well as the laser powers and detunings, the platform may be used as a photon-number-sensitive filter of optical pulses by selecting transparency for specific photon number states in the pair-interaction energy landscape.

% References
\section*{References}
\bibliographystyle{iopart-num}
\bibliography{flashbib}

\providecommand{\newblock}{}
\begin{thebibliography}{10}
\expandafter\ifx\csname url\endcsname\relax
  \def\url#1{{\tt #1}}\fi
\expandafter\ifx\csname urlprefix\endcsname\relax\def\urlprefix{URL }\fi
\providecommand{\eprint}[2][]{\url{#2}}

\bibitem{Moh2007}
Mohapatra A~K, Jackson T~R and Adams C~S 2007 {\em Physical Review Letters\/}
  {\bf 98} 113003

\bibitem{Pri2010}
Pritchard J~D, Maxwell D, Gauguet A, Weatherill K~J, Jones M~P~A and Adams C~S
  2010 {\em Physical Review Letters\/} {\bf 105} 193603

\bibitem{Peyronel2012}
Peyronel T, Firstenberg O, Liang Q~Y, Hofferberth S, Gorshkov A~V, Pohl T,
  Lukin M~D and Vuleti{\'{c}} V 2012 {\em Nature\/} {\bf 488} 57--60

\bibitem{Tiarks2019}
Tiarks D, Schmidt-Eberle S, Stolz T, Rempe G and D{\"u}rr S 2019 {\em Nature
  Physics\/} {\bf 15} 124

\bibitem{Dudin2012}
Dudin Y~O and Kuzmich A 2012 {\em Science\/} {\bf 336} 887--889

\bibitem{Ripka2018}
Ripka F, K{\"{u}}bler H, L{\"{o}}w R and Pfau T 2018 {\em Science\/} {\bf 362}
  446--449

\bibitem{Gorshkov2011}
Gorshkov A~V, Otterbach J, Fleischhauer M, Pohl T and Lukin M~D 2011 {\em
  Physical Review Letters\/} {\bf 107} 1--5

\bibitem{Gorshkov2013}
Gorshkov A~V, Nath R and Pohl T 2013 {\em Physical Review Letters\/} {\bf 110}
  153601

\bibitem{Khazali15}
Khazali M, Heshami K and Simon C 2015 {\em Physical Review A\/} {\bf 91}
  030301(R)

\bibitem{Sevincli2011}
S~Sevincli N~H, Ates C and Pohl T 2011 {\em Physical Review Letters\/} {\bf
  107} 153001

\bibitem{gorniaczyk16}
Gorniaczyk H, Tresp C, Bienias P, Paris-Mandoki A, Li W, Mirgorodskiy I,
  B{\"u}chler H~P, Lesanovsky I and Hofferberth S 2016 {\em Nature
  Communications\/} {\bf 7} 12480

\bibitem{petrosyan11}
Petrosyan D, Otterbach J and Fleischhauer M 2011 {\em Physical Review
  Letters\/} {\bf 107} 213601

\bibitem{Li14}
Li W, Viscor D, Hofferberth S and Lesanovsky I 2014 {\em Physical Review
  Letters\/} {\bf 112} 243601

\bibitem{Firstenberg2013}
Firstenberg O, Peyronel T, Liang Q~Y, Gorshkov A~V, Lukin M~D and Vuleti{\'{c}}
  V 2013 {\em Nature\/} {\bf 502} 71--75

\bibitem{Jenkins2016}
Jenkins S~D, Ruostekoski J, Javanainen J, Jennewein S, Bourgain R, Pellegrino
  J, Sortais Y~R~P and Browaeys A 2016 {\em Physical Review A\/} {\bf 94}
  023842

\bibitem{Roof2016}
Roof S~J, Kemp K, Havey M~D and Sokolov I~M 2016 {\em Physical Review
  Letters\/} {\bf 117} 073003

\bibitem{Guerin2016}
Ara{\'{u}}jo M~O, Kre{\v{s}}i{\'{c}} I, Kaiser R and Guerin W 2016 {\em
  Physical Review Letters\/} {\bf 117} 073002

\bibitem{Bradac2017}
Bradac C, Johnsson M~T, van Breugel M, Baragiola B~Q, Martin R, Juan M~L,
  Brennen G~K and Volz T 2017 {\em Nature Communications\/} {\bf 8} 1205

\bibitem{Facchinetti2016a}
Facchinetti G, Jenkins S~D and Ruostekoski J 2016 {\em Physical Review
  Letters\/} {\bf 117} 243601

\bibitem{Guerin2017}
Guerin W and Kaiser R 2017 {\em Physical Review A\/} {\bf 95} 053865

\bibitem{Chen2010a}
Chen J~F, Loy M~M, Wong G~K and Du S 2010 {\em Journal of Optics\/} {\bf 12}
  ISSN 20408978

\bibitem{Toyoda1997}
Toyoda K, Takahashi Y, Ishikawa K and Yabuzaki T 1997 {\em Physical Review A\/}
  {\bf 56} 1564

\bibitem{Bettles2018}
Bettles R~J, Ilieva T, Busche H, Huillery P, Ball S~W, Spong N~L~R and Adams
  C~S 2018   1--7 (\textit{Preprint} \eprint{1808.08415})

\bibitem{Chen2010}
Chen J~F, Wang S, Wei D, Loy M~M~T, Wong G~K~L and Du S 2010 {\em Physical
  Review A\/} {\bf 81} 033844

\bibitem{Busche2016}
Busche H, Ball S~W and Huillery P 2016 {\em European Physical Journal: Special
  Topics\/} {\bf 225} 2839--2861

\bibitem{Weber2015}
Weber T~M, H{\"{o}}ning M, Niederpr{\"{u}}m T, Manthey T, Thomas O, Guarrera V,
  Fleischhauer M, Barontini G and Ott H 2015 {\em Nature Physics\/} {\bf 11}
  157--161

\bibitem{Zeuthen2017a}
Zeuthen E, Gullans M~J, Maghrebi M~F and Gorshkov A~V 2017 {\em Physical Review
  Letters\/} {\bf 119} 043602

\bibitem{zhang_bai_huang_2018}
Zhang Q, Bai Z and Huang G 2018 {\em Physical Review A\/} {\bf 97} 043821

\bibitem{Moehl2019}
M\"ohl C 2019 {\em Master thesis\/}

\bibitem{Steck}
Steck D~A 2001

\bibitem{Hau1999}
Hau L~V, Harris S~E, Dutton Z and Behroozi C~H 1999 {\em Nature\/} {\bf 397}
  594--598

\bibitem{Maxwell2013}
Maxwell D, Szwer D~J, Paredes-Barato D, Busche H, Pritchard J~D, Gauguet A,
  Weatherill K~J, Jones M~P~A and Adams C~S 2013 {\em Physical Review
  Letters\/} {\bf 110} 103001

\bibitem{Bariani2012}
Bariani F, Dudin Y~O, Kennedy T~A~B and Kuzmich A 2012 {\em Physical Review
  Letters\/} {\bf 108} 030501

\bibitem{Yan2019}
Yan D, Wang B, Bai Z and Li W 2019   1--8 (\textit{Preprint}
  \eprint{1902.07492}) \urlprefix\url{http://arxiv.org/abs/1902.07492}

\end{thebibliography}

\end{document}